# Electroporation-mediated Metformin for effective anticancer treatment of triple-negative breast cancer cells

Praveen Sahu[1], Ignacio G. Camarillo[2,3], Pragatheiswar Giri[1] and Raji Sundararajan[1]
[1] School of Engineering Technology, Purdue University, Indiana, USA
[2] Department of Biological Sciences, Purdue University, Indiana, USA
[3] Purdue University Center for Cancer Research, Indiana, USA
phone: (+1) 765-772-0929
e-mail: psahu@purdue.edu

*Abstract*— In this research, we investigated the efficacy of Metformin, the most commonly administered type-2 diabetes drug for triple negative breast cancer (TNBC) treatment, due to its various anticancer properties. It is a plant-based bio-compound, synthesized as a novel biguanide, called dimethyl biguanide or metformin. One of the ways it operates is by hindering electron transport chain-complex I, in mitochondria, which causes a drop-in energy (ATP) generation. This eventually builds energetic stress and a decline in energy. Therefore, the natural cellular processes and proliferating tumor cells are obstructed. Here, we used electroporation, where, the MDA-MB-231, human TNBC cells were subjected to high intensity, short-duration electrical pulses (EP) in the presence of Metformin. The cell viability results indicate lower cell viability of 43.45% as compared to 85.20% with drug alone at 5mM concentration. This indicates that Metformin, the most common diabetes drug could also be explored for cancer treatment.

## I. INTRODUCTION

Triple-Negative Breast Cancer (TNBC) is an aberrant type of breast cancer subset, with the absence of traditional biomarkers *i.e.,* Estrogen and progesterone receptors, and shows no upregulation in HER2 protein. It is the most aggressive subset of all breast cancer types. Hence, making the diagnosis and treatment of TNBC is indeed challenging, and novel therapies are needed. In this light, various attempts have been made to combine the application of Electrical pulse with various drug combinations for enhanced modality [1]–[3]. In this research, we explored the anticancer properties of Metformin, the most commonly used Type-2 diabetes drug.

Metformin has shown various anticancer properties[4], [5]. Subsequently, with a better understanding of its working mechanism, which includes multiple factors, such as inhibition of specific signaling pathways, upregulation of molecules involved in apoptosis, oxidative stress, *etc*.[6]-[8], we selected Metformin to investigate its effect on a human

TNBC cell line. TNBC was selected due to its aggressive nature, with an unmet need for novel therapies.

Electroporation was used to enhance the uptake of the drug molecules, and the effects on cell viability and change in oxidative stress were investigated using MDA-MB-231, the human triple-negative breast cancer cell line.

## II. MATERIAL AND METHODS

### A. TNBC Cell Line

MDA-MB-231 (ATCC® HTB-26™) is an epithelial-like cancerous cell taken from tissue of triple-negative breast cancer metastatic site. The single layer of this cell was cultured on a T-75 flask under specific conditions and media mentioned in the next section.

### B. Culture Media and Growth Conditions

Commercial Gibco™ Dulbecco's Modified Eagle Medium (DMEM) was used to culture the MDA-MB-231 cells line with the combination of 10% FBS and 1% Penicillin-Streptomycin. All the items were procured from Thermofisher Scientific™. We used the standard incubated condition of 37°C at 80-85% humidity in the presence of 5% $CO_2$. The morphology of the untreated cell before the experiment is shown in Fig. 1.

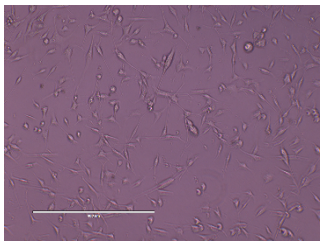

Fig. 1. MDA-MB-231 untreated cell morphology at 10X optical zoom.

### C. Metformin

Metformin hydrochloride (1,1-Dimethylbiguanide hydrochloride) was procured from Sigma-Aldrich™. It was dissolved in DI (De-ionized water) to achieve the desired concentration of 1mM and 5mM for comparative study. The chemical structure has been shown in Fig. 2.

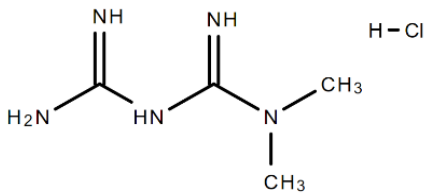

Fig. 2. The chemical structure of commercially available Metformin hydrochloride (Sigma Aldrich™)

### D. Electroporator

We used BTX ECM 830 electroporator by *Genetronics Inc. San Diego, CA, USA.* This electroporator can deliver electrical field strengths in the range of 200V/cm to 5000V/cm. Throughout our experiments, we applied 8-pulses of 1000V/cm at 100μs, at 1s intervals. The electroporation *in-vitro* study was performed using 4mm gap sterile electroporation cuvettes, which hold 600μl of MDA-MB-231 cell suspension at $1 \times 10^6$ cells/ml concentration.

### E. Cell Viability

We used advanced RealTime-Glo™ MT Cell Viability Assay by Promega which is capable of real-time monitoring [9]. For the samples without EP treatment, 20μl cell suspension (20,000 cells/well) was directly dispensed with appropriate drug concentration into a 96-well plate with 55μl of cell media and 25μl of MT assay reagent. A similar exercise was done for the remaining samples but after performing the electroporation at the given EP parameter. Next, the cells were incubated and after 24h and luminescence (Lum) reading was taken using Synergy HTX Multi-Mode microplate Reader. The final percentage viability was reported by normalized to the control Lum value.

$$\% \; Cell \; Viability = \frac{Sample \; Luminescence \; Value}{Control \; Luminescence \; Value} \; x \; 100 \qquad (1)$$

### F. Reactive Oxygen Species (ROS) Study

We performed ROS-Glo™ $H_2O_2$ Assay to measure the level of hydrogen peroxide ($H_2O_2$), a reactive oxygen species (ROS)[10]. It allows us to measure the oxidative stress in the treated cell alongside the cell viability. All the samples were dispensed in a 96-well plate with 20000 cells/wells with 80μl of cell media and incubated. After 18h of incubation, 20μl of $H_2O_2$ substrate was introduced allowed to further incubate for 6h. Next, 100μl luciferase detection reagent followed by 20min incubation time and Luminescence (RLU) using Synergy HTX Multi-Mode microplate Reader.

### G. Statistical Analysis

The cell viability and ROS study were subjected to Repeated Measure Analysis of Variance ANOVA for statistical significance, followed by Tukey's test for multiple comparisons[11], [12]. The significance level was set to α=0.05.

We have a total of five treatments (k=5) and each treatment is assigned with a single letter (alphabet say 'A', 'B' etc.). But before we assigned a letter to any treatment we calculate the Critical Value (CV) in Tukey's test given by the following equation [13]:

$$Critical \; Value \; (CV) = q * sqrt\left(\frac{MS}{n}\right) \qquad (2)$$

Here, q-value is picked up from standard q-score Tukey's table corresponding to the specific number of treatment (k) and degree of freedom (df) within treatments and MS is Mean Square within the treatments from One-way ANOVA analysis. Here, n=3, since each of the sample's treatments was performed in triplicate. Now, we assign the same alphabets to those treatments, which do not significantly different, *i.e.* the difference between the treatment is less than that of Critical Value (CV). Otherwise, they are represented with different alphabets (p<0.05). The statistical analysis was done using R-studio. The data was represented in the form of Mean ± Standard Error (μ ± SE).

### III. RESULT AND DISCUSSION

#### A. Cell Viability Study

Fig 3 shows the cell viabilities for the all the treatments, *i.e.* No treatment (Control), 1mM Metformin, 5mM Metformin, 1mM Metformin + EP and 5mM Metformin + EP at 24h. The cell viabilities were normalized to 100% *w.r.t* control at 24h.

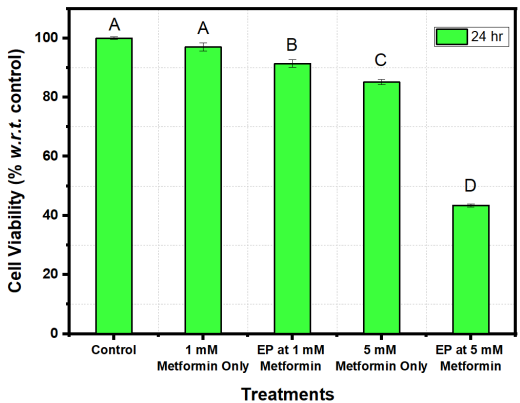

Fig. 3. Cell Viability at 24 h for 1mM and 5 mM metformin with and without Electroporation. The error bars represent the Standard Error (SE) and different letters depict a significant difference p<0.05 from Tukey's Test.

We opted for Tukey's test to figure out the significance among the various treatments and the summarization from Table 1 ANOVA result was used to calculate the critical value as 4.59 as shown in Table 2. This indicated that we cannot consider two treatments significantly different unless and until the difference is greater than CV=4.59. Hence, at 1 mM Metformin there is only a minimal reduction of 3.03 % which is also insignificant as per the Tukey's test and hence assigned the same letter 'A' as compared to no treatment control. Whereas treatment at 1 mM along with EP shows cell viability of 91.41% and further cell viability of 5mM Metformin with EP is 43.45% compared to 85.2% with 5mM Metformin alone. These are assigned different letters as the difference is greater than the critical values.

TABLE 1: SUMMARY OF ANOVA ANALYSIS ON THE CELL VIABILITY TREATMENT

| Source of variation | Degree of Freedom (df) | Mean Square (MS) | P-Values |
|---|---|---|---|
| Between treatments | 4 | 1592.65 | 1.18832E-11 |
| Within treatments | 10 | 2.92 | |

TABLE 2: TUKEY TEST PARAMETERS AND ANALYSIS RESULTS

| Parameter | Values |
|---|---|
| Total Treatment Group (k) | 5 |
| Degree of Freedom (df) | 10 |
| q-value for Tukey table | 4.65 |
| Critical value (CV) | 4.59 |
| Significance Value ($\alpha$) | 0.05 |

B. *Reactive Oxygen Species (ROS) Study*

Fig 4 shows the parallel study done on the treatments to quantify the oxidative stress in TNBC cells. The critical value for ROS significance among various treatments from Tukey's Test (Table 3) and ANOVA analysis (Table 4) was found to be 71.13. It means that for any two treatments to be significantly different the change in ROS has to be more than 71.13.

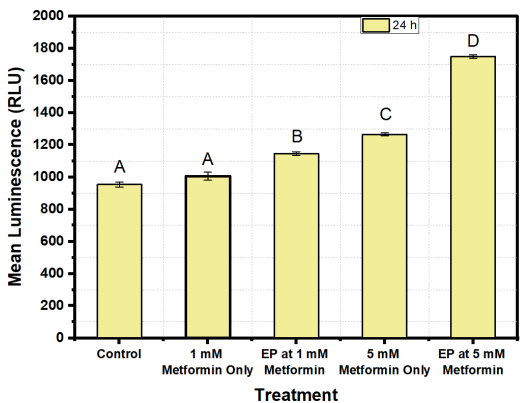

Fig. 4. Oxidative stress represented by the change in $H_2O_2$ reactive oxygen species (ROS) in MDA-MB-231 at 24 h for 1mM and 5mM metformin with and without Electroporation including no treatment (Control). The error bars represent the standard error and different letters depict a significant difference p<0.05 from Tukey's Test.

TABLE 3: TUKEY TEST PARAMETERS AND ANALYSIS RESULTS

| Parameter | Values |
|---|---|
| Total Treatment Group (k) | 5 |
| Degree of Freedom (df) | 10 |
| q-value for Tukey table | 4.65 |
| Critical value (CV) | 71.13 |
| Significance Value ($\alpha$) | 0.05 |

TABLE 4: SUMMARY OF ANOVA ANALYSIS ON THE ROS TREATMENT

| Source of variation | Degree of Freedom (df) | Mean Square (MS) | P-Values |
|---|---|---|---|
| Between treatments | 4 | 303131.16 | 3.77186E-11 |
| Within treatments | 10 | 701.93 | |

The Mean Lum was 954 for Control, which insignificantly increased to 1005 at 1mM Metformin, hence, the same alphabet 'A'. Along with EP, it increases to 1147 and further to 1265 at 5mM Metformin as depicted by the alphabet 'B' and 'C' respectively. The $H_2O_2$ reactive oxygen species (ROS) levels were found to be maximum mean Lum of 1749 in the case of 5mM Metformin + EP trigger which suggests a significantly higher change in oxidative stress [14] in TNBC cells at 24h leading to higher cell death as shown in Fig. 4.

## IV. CONCLUSION

MDA-MB-231 cells were treated with 8-pulses of 1000V/cm at 100μsec,1s intervals for 1mM and 5mM Metformin. The viability results indicate that EP + 5mM Metformin has cell viability is as low as 43.45% as compared to 82.50% for its counterpart *i.e.,* Metformin alone. The same is corroborated by the higher ROS oxidative stress level of 1910 Lum for EP along with 5mM Metformin. These indicate the potency of the electroporation mediated Metformin approach could present a valuable alternative to target TNBC cells.

*Proc. 2023 Annual Meeting of the Electrostatics Society of America*